\documentclass[prl,twocolumn,showpacs,preprintnumbers,amsmath,amssymb]{revtex4}

\usepackage{graphicx}
\usepackage{dcolumn}
\usepackage{bm}

\begin{document}

\textbf{Comment on ``Non-monotonicity in the Quantum-Classical Transition: Chaos Induced by Quantum Effects"} 

In a recent Letter~\cite{Kapulkin08}, Kapulkin and Pattanayak presented results regarding the chaotic behavior of the damped Duffing oscillator as it undergoes the transition from quantum to classical dynamics, this transition being induced by a model of continuous observation referred to as quantum-state diffusion~\cite{Percival98}. They present evidence that a Duffing oscillator, sufficiently damped so that it is {\em not} classically chaotic, becomes chaotic in the transition region. If true, this would be a striking result.  However, Kapulkin and Pattanayak did not calculate the Lyapunov exponent for the system, usually regarded as the litmus-test of chaos. Here we perform this calculation, which throws considerable doubt upon the conclusions in~\cite{Kapulkin08}. 

Since dynamical systems become very noisy as they pass through the transition, to calculate a Lyapunov exponent at all, one must separate the sensitivity (chaos) induced by the deterministic dynamics from the unpredictability due to the noise. This can be achieved by comparing the evolutions of the system for two nearby points in phase-space, using the {\em same} noise realization for both. If the separation of these trajectories, given by the usual distance in Hilbert space, is $\Delta(t)$, then the Lyapunov exponent is $\lambda =  \lim_{t\rightarrow\infty} \left\{ \lim_{\Delta (0)\rightarrow 0} \left( \ln [ \Delta(t)/\Delta(0)]  /t  \right) \right\}$. 
 
To calculate $\lambda$ numerically, $\Delta (t)$ must remain sufficiently small during the entire evolution that its dynamics is described to high accuracy by the  dynamics of the system linearized about either of the trajectories. This can be realized by periodically re-scaling one of the trajectories towards the other along a line in the state-space of the system (which is in our case is simply the Hilbert space). This elegant procedure, a version of the Wolf method~\cite{Wolf85}, works because the linearized dynamics is unaffected by the size of $\Delta(t)$.  We calculate the Lyapunov exponent by evolving a pair of trajectories for $\sim 7000$ periods of the driving. To increase the accuracy we averaged the resulting exponent over 124 pairs, each with a different noise realization. For a given pair, the result obtained for the Lyapunov exponent, after evolving for a time $t$, initially falls as $1/t$, and then flattens out as it approaches and stabilizes at the true value. We report either this stabilized value, with the error determined by the spread over the different noise realizations (this error reduces as $t$ increases),  or, if the estimated value never flattens out, we report the resulting upper bound on the Lyapunov exponent. 

The Hamiltonian for the Duffing oscillator used in~\cite{Kapulkin08} is given by 
\begin{equation}
H = \hbar \omega \left[ \frac{P^2}{2} - \frac{X^2}{2} + \beta^2  \frac{X^4}{4}  - \frac{g}{\beta} X\cos(\Omega t) \right] , 
\end{equation}
\begin{figure}[t]
\leavevmode\includegraphics[width=1\hsize]{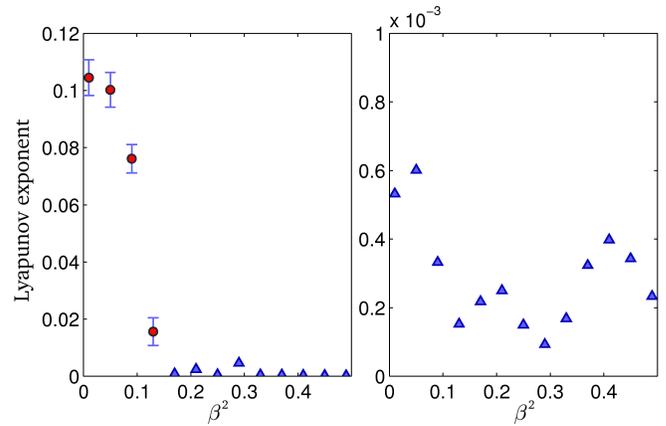} 
\caption{(Color online) The Lyapunov exponent for the damped quantum Duffing oscillator as a function of the inverse system size, $\beta^2$. For $\beta^2=0.01$ the system is essentially classical. Red circles give the Lyapunov exponent, $\lambda$, while blue triangles give upper bounds on $\lambda$. (a) Small damping so that the oscillator is chaotic in the classical regime; (b) larger damping so that the classical oscillator has no chaos. } 
\label{fig1} 
\end{figure} 

\hspace{-3.2ex} where $X$ and $P$ are position and momentum scaled so that $[X,P] = i$, and $\omega$ is a frequency. Since the initial factor of $\hbar$ does not affect the dynamics, the size (action) of the system with respect to $\hbar$ is entirely captured by the parameter $\beta = \sqrt{\hbar/(m\omega l^2)}$, where $m$ and $l$ are, respectively, a mass and length scale. Damping given by $\dot{P} = -\Gamma P$ is induced by the quantum-state diffusion model~\cite{Percival98}. Scaling time by $1/\omega$, Kapulkin and Pattanayak use the parameters $\Omega=1, g = 0.3$ and either $\Gamma = 0.125$ (classically chaotic) or $\Gamma = 0.3$ (classically unchaotic). Our results for these two cases are shown in Fig.~1. For the classically chaotic case the Lyapunov exponent decays as expected as $\hbar$ increases, and the system moves into the quantum regime. However, when the system is not classically chaotic, the Lyapunov exponent remains zero throughout.  At least, our results place an upper bound on this exponent of $10^{-3}$. This throws considerable doubt on the conclusion in~\cite{Kapulkin08} that chaos emerges in the transition region for this system. 

\vspace{2ex}
\hspace{-3ex} Justin Finn, Kurt Jacobs and Bala Sundaram

Department of Physics, 

University of Massachusetts at Boston, 

100 Morrissey Blvd, 

Boston, MA 02125, USA


\end{document}